\documentclass[runningheads]{llncs}

\usepackage{placeins}
\usepackage[T1]{fontenc}
\usepackage{graphicx}
\graphicspath{{Figures/}{./}}
\usepackage{wrapfig}

\usepackage{amsmath}
\usepackage{xspace}
\usepackage{pifont}
\usepackage{enumitem}
\usepackage{times}
\usepackage{hyperref}

\usepackage{xurl}

\usepackage{tikz}        
\usepackage{pgfplots}
\pgfplotsset{compat=1.18}

\usepackage{booktabs}
\usepackage{threeparttable}
\usepackage{array}
\usepackage{tabularx}

\newcommand{\cmark}{\ding{51}}      
\newcommand{\pmark}{\(\circ\)} 

\newcommand{\nmark}{--}             

\newcommand{\BfPara}[1]{\noindent\textbf{#1.}\xspace}
\newcolumntype{Y}{>{\raggedright\arraybackslash}X}

\newcommand{\ours}{{RAG-PIBench}\xspace}

\begin{document}

\title{\ours{}: A Leakage-Aware Benchmark for Prompt-Injection Detection in Trustworthy RAG Systems}

\titlerunning{Prompt-Injection Detection for Trustworthy RAG Systems}

\author{
Niveen O. Jaffal\inst{1} \and
Ahmet Yuksel\inst{2} \and
David Mohaisen\inst{3}
}

\authorrunning{N. O. Jaffal et al.}

\institute{
Birzeit University, Birzeit, Palestine\\
\email{njaffal@birzeit.edu}
\and
Independent Researcher\\
\email{asyuksel5@gmail.com}
\and
University of Central Florida, Orlando, FL, USA\\
\email{mohaisen@ucf.edu}
}



\maketitle

\begin{abstract}
Retrieval-Augmented Generation (RAG) systems are vulnerable to prompt-injection attacks embedded in retrieved content. We introduce \ours{}, a benchmark for RAG-style prompt-injection detection containing 4,876 contextual examples across frozen train, validation, and protected-test splits. Using a leakage-aware construction pipeline and strict evaluation protocol, we compare keyword-based, semantic-reference, TF-IDF, and transformer-based detectors. DistilBERT achieves the best protected-test performance (\(F_1=0.896\), PR-AUC \(=0.968\)), while TF-IDF SVM and logistic regression remain competitive. Our results demonstrate the value of leakage-aware benchmark design and strong sparse baselines for reliable prompt-injection detection in RAG systems.

\keywords{Prompt injection \and Retrieval-Augmented Generation \and RAG security \and Large language models \and Benchmark construction \and Prompt-injection detection}
\end{abstract}

\section{Introduction \& Problem Statement}
\label{sec:introduction_problem_statement}

Retrieval-Augmented Generation (RAG) has become a widely adopted approach for improving large language models (LLMs) by incorporating external documents into the generation context~\cite{watsonbenjamin2025contextfusion,oro2024evaluating,ogdu2025adaptive}. While RAG can improve factual grounding and domain adaptation~\cite{rakin2024leveraging}, it also expands the attack surface of LLM-based systems. In particular, prompt-injection attacks can be embedded within retrieved content, causing the model to follow adversarial instructions rather than the intended system objective~\cite{shrivastav2025effective,destefano2024rag}.

Detecting prompt injection in RAG contexts is challenging because benign documents may naturally contain instructions, code snippets, examples, or task descriptions. Moreover, malicious instructions can be paraphrased, blended into otherwise legitimate content, or designed to resemble benign contextual information. As a result, simple keyword-based defenses often fail, while more sophisticated approaches may over-flag benign content or rely on dataset artifacts rather than robust semantic signals~\cite{geng2026prompt,hadiprakoso2025adaptive,guan2025privacy}.

A variety of detection approaches have been proposed, ranging from heuristic filters and semantic-reference methods to sparse-feature classifiers and transformer-based models~\cite{gallifant2025sparse,latibari2024transformers}. However, comparisons across these detector families are often complicated by inconsistent evaluation protocols, potential leakage between training and testing data, and the absence of strong lightweight baselines.

\BfPara{Problem Statement}
Despite growing interest in prompt-injection defenses, there remains a need for leakage-aware benchmarks that evaluate whether detectors can reliably distinguish benign contextual instructions from malicious instruction-override attempts in realistic RAG inputs. Existing evaluations may overestimate performance due to lexical shortcuts, repeated templates, or insufficient separation between dataset construction and testing. This work addresses the problem of constructing and evaluating a reproducible RAG-style prompt-injection benchmark that supports fair comparison across heuristic, semantic, sparse-feature, and transformer-based detectors.

\BfPara{Research Questions}
This study is guided by four research questions: \textbf{RQ1.} How effectively do different detector families distinguish benign RAG-style contextual inputs from malicious prompt-injection under a frozen protected-test evaluation protocol? \textbf{RQ2.} Do compact transformer classifiers provide measurable gains over strong baseline detectors, or can sparse lexical and character-level models remain competitive? \textbf{RQ3.} What precision--recall trade-offs and failure modes emerge across keyword heuristics, semantic-reference methods, TF-IDF classifiers, and transformer models? \textbf{RQ4.} How can dataset construction, split isolation, diagnostic separation, and artifact registration reduce leakage and improve the reliability of prompt-injection detector evaluation?

\BfPara{Contributions}
To address these questions, this paper makes five contributions: \ding{182} \BfPara{\ours{} Benchmark} We introduce \ours{}, a balanced benchmark for RAG-style prompt-injection detection containing 4,876 rendered contextual examples across frozen train, validation, and protected-test splits. \ding{183} \BfPara{Leakage-Aware Construction Pipeline} We develop a dataset-construction pipeline incorporating source registration, provenance tracing, quality control, style-matched rendering, shortcut auditing, and frozen split generation to reduce shortcut learning and improve auditability. \ding{184} \BfPara{Protected-Test Evaluation Protocol} We implement a strict train/validation/protected-test protocol that isolates model training, selection, and final evaluation while keeping diagnostic analyses separate from tuning decisions. \ding{185} \BfPara{Systematic Baseline Evaluation} We compare keyword rules, semantic-reference methods, TF-IDF LR, TF-IDF SVM, TF-IDF RF, DistilBERT, and DeBERTa-v3 to assess whether transformer models provide meaningful gains over strong lightweight baselines. \ding{186} \BfPara{Transparent Findings and Failure Modes} We report both strengths and weaknesses of the evaluated detectors, showing that DistilBERT achieves the best protected-test performance while TF-IDF SVM and TF-IDF LR remain highly competitive, and identifying key failure modes across detector families.

\BfPara{Positioning} RAG-PIBench extends our broader line of work on trustworthy and secure LLMs, which has examined LLM attack surfaces and defenses, security of generated code, attribution of LLM-generated artifacts, and systematic evaluation of LLMs in security applications \cite{jaffal2025large,kharma2026security,choi2025attributing,lin2025large}. Here, we focus specifically on indirect prompt injection in RAG settings and on leakage-aware evaluation of prompt-injection detectors.

\section{Related Work}
\label{sec:related_work}

Research on prompt-injection security spans benchmark construction, detector design, and RAG-specific defenses. We focus on work most closely related to leakage-aware evaluation, indirect prompt injection, and detection in retrieval-augmented settings.

\BfPara{Prompt-Injection Benchmarks}
\label{subsec:prompt_injection_benchmarks} 
Recent benchmarks have moved beyond simple prompt-level attacks toward indirect, application-oriented, and agent-based threat models. PromptShield~\cite{jacob2025promptshield} studies practical prompt-injection detection under realistic false-positive constraints, while PromptSleuth~\cite{wang2025promptsleuth} investigates robustness under paraphrased and distribution-shifted attacks. BIPIA~\cite{yi2025bipia} focuses on indirect prompt injection through retrieved content in RAG workflows, whereas INJECAGENT~\cite{zhan2024injecagent} and WAInjectBench~\cite{liu2025wainjectbench} extend evaluation to tool-integrated and web-agent environments. Collectively, these benchmarks demonstrate the importance of contextualized attacks and distribution-aware evaluation, but they provide limited emphasis on leakage-aware detector assessment and protected-test benchmarking.

\BfPara{Prompt-Injection Defenses in RAG}
\label{subsec:prompt_injection_defenses_rag} 
Defenses against prompt injection range from lightweight filters and semantic detectors to multi-stage RAG frameworks. Prior work has explored adaptive defense pipelines~\cite{hadiprakoso2025adaptive}, post-generation validation using auxiliary LLMs~\cite{helbling2023llm}, retrieval-aware defenses such as TrustRAG~\cite{zhou2025trustrag} and SeCon-RAG~\cite{si2025seconrag}, and hybrid architectures combining rules, classifiers, and companion models~\cite{kokkula2024palisade}. While these approaches improve robustness, their evaluation protocols vary considerably, making direct comparison difficult. Reliable detector assessment therefore requires controlled benchmarks with strict separation between training, validation, and final testing.

\BfPara{Positioning of This Work}
\label{subsec:positioning} 
\ours{} complements existing benchmarks and defenses by focusing on leakage-aware binary detection of prompt injection in RAG-style contextual inputs. Unlike prior studies that primarily measure attack success or evaluate a single detector family, we compare heuristic, semantic, sparse-feature, and transformer-based detectors under a frozen train/validation/protected-test protocol. The benchmark further incorporates provenance tracking, quality control, contextual rendering, shortcut auditing, and strict split isolation to support reproducible evaluation.

Table~\ref{tab:related_work_positioning} summarizes the positioning of \ours{} relative to representative prompt-injection benchmarks and trustworthy-RAG defenses. The comparison highlights three distinguishing characteristics of our study: (i) leakage-aware benchmark construction, (ii) evaluation against strong lightweight baselines in addition to transformer models, and (iii) explicit reporting of detector failure modes alongside successful results.

\begin{table}[!t]
\centering
\caption{Comparison of representative prompt-injection benchmarks, detection frameworks, and trustworthy-RAG defenses. Columns indicate whether the work explicitly addresses indirect prompt injection (IPI), retrieval-augmented generation (RAG), text and image modalities, leakage-aware or protected evaluation protocols, detector baseline comparisons, and failure-mode analysis. Modalities are text (T) or image (I). The attributes of the work are leakage (L), baselines (B), and failures (F)}
\label{tab:related_work_positioning}
\scriptsize
\setlength{\tabcolsep}{3pt}
\renewcommand{\arraystretch}{1.12}

\begin{tabular}{@{}p{2.3cm}c p{2.1cm} c c cc c c c c@{}}
\toprule
\textbf{Work} &
\textbf{Yr.} &
\textbf{Setting} &
\textbf{IPI} &
\textbf{RAG} &
\multicolumn{2}{c}{\textbf{Modality}} &
\textbf{Goal} &
\multicolumn{3}{c}{\textbf{Attributes}} \\
\cmidrule(lr){6-7}
& & & & & \textbf{T} & \textbf{I} & & L & B & F\\
\midrule

Palisade~\cite{kokkula2024palisade}
& 2024 & PI detection
& \pmark & \nmark & \cmark & \nmark
& Filter/classifier defense
& \nmark & \pmark & \pmark \\

INJECAGENT~\cite{zhan2024injecagent}
& 2024 & Tool-agent PI
& \cmark & \nmark & \cmark & \nmark
& Agent vulnerability
& \nmark & \nmark & \pmark \\

PromptShield~\cite{jacob2025promptshield}
& 2025 & PI detection
& \cmark & \nmark & \cmark & \nmark
& PI detection
& \nmark & \cmark & \pmark \\

PromptSleuth~\cite{wang2025promptsleuth}
& 2025 & Semantic PI det.
& \cmark & \nmark & \cmark & \nmark
& Semantic detection
& \pmark & \cmark & \pmark \\

BIPIA~\cite{yi2025bipia}
& 2025 & External-content PI
& \cmark & \cmark & \cmark & \nmark
& Vulnerability and defense
& \pmark & \pmark & \pmark \\

WAInjectBench~\cite{liu2025wainjectbench}
& 2025 & Web-agent PI
& \cmark & \nmark & \cmark & \cmark
& Web-agent detection
& \pmark & \cmark & \cmark \\

TrustRAG~\cite{zhou2025trustrag}
& 2025 & Trustworthy RAG
& \pmark & \cmark & \cmark & \nmark
& RAG defense
& \nmark & \cmark & \pmark \\

SeCon-RAG~\cite{si2025seconrag}
& 2025 & RAG conflict filter
& \nmark & \cmark & \cmark & \nmark
& RAG defense
& \nmark & \cmark & \pmark \\

\midrule

\textbf{\ours{}}
& \textbf{2026}
& \textbf{RAG-style contextual PI detection}
& \textbf{\cmark}
& \textbf{\cmark}
& \textbf{\cmark}
& \textbf{\nmark}
& \textbf{Leakage-aware PI detection}
& \textbf{\cmark}
& \textbf{\cmark}
& \textbf{\cmark} \\

\bottomrule
\end{tabular}
\end{table}

\section{Methodology}
\label{sec:methodology}

This section describes the construction of \ours{} and the evaluation protocol used to assess prompt-injection detectors in a controlled, leakage-aware setting. The methodology supports fair comparison across heuristic, sparse-feature, and transformer-based models while maintaining strict separation between training, validation, protected testing, and diagnostic analyses.

\subsection{Threat Model}
\label{subsec:threat_model}

We consider an indirect prompt-injection threat model for RAG-style systems in which an attacker controls or influences external textual content that may be retrieved and incorporated into the model context. The attacker does not control the user query or system prompt; instead, malicious instructions are embedded within ordinary retrieved content. The detector operates before generation on the rendered RAG-style input and performs binary classification. Malicious examples are defined as inputs that attempt to override, bypass, redirect, or manipulate the assistant's intended instruction-following behavior, while benign examples contain contextual content without such intent.

Our focus is prompt-injection detection rather than end-to-end attack execution. Accordingly, we do not evaluate downstream generation outcomes, tool misuse, data exfiltration, retrieval poisoning, or post-detection mitigation strategies. The benchmark is restricted to English textual inputs and evaluated under a frozen train, validation, and protected-test protocol.

\subsection{Task Definition}
\label{subsec:task_definition}

We formulate prompt-injection detection in RAG-style contexts as a binary classification task. Each example consists of a rendered textual input \(x_i\) containing a host context and an embedded content segment. The objective is to determine whether the embedded content is benign or contains a malicious instruction intended to alter the behavior of a downstream assistant. Formally, each instance is assigned a label \(y_i \in \{0,1\}\), where \(0\) denotes benign content and \(1\) denotes prompt injection. A detector learns a mapping \(f(x_i) \rightarrow y_i\). Performance is evaluated not only by overall classification effectiveness but also by the ability to balance detection of malicious content against false positives on benign instruction-like text. This formulation captures a key challenge in RAG security: benign retrieved content may naturally contain instructions, examples, code snippets, or task descriptions, requiring detectors to distinguish legitimate instructional content from adversarial instruction-override attempts.

\subsection{Overview of the \ours{} Construction Pipeline}
\label{subsec:construction_overview}

\begin{figure*}[t]
    \centering
    \includegraphics[width=0.98\textwidth]{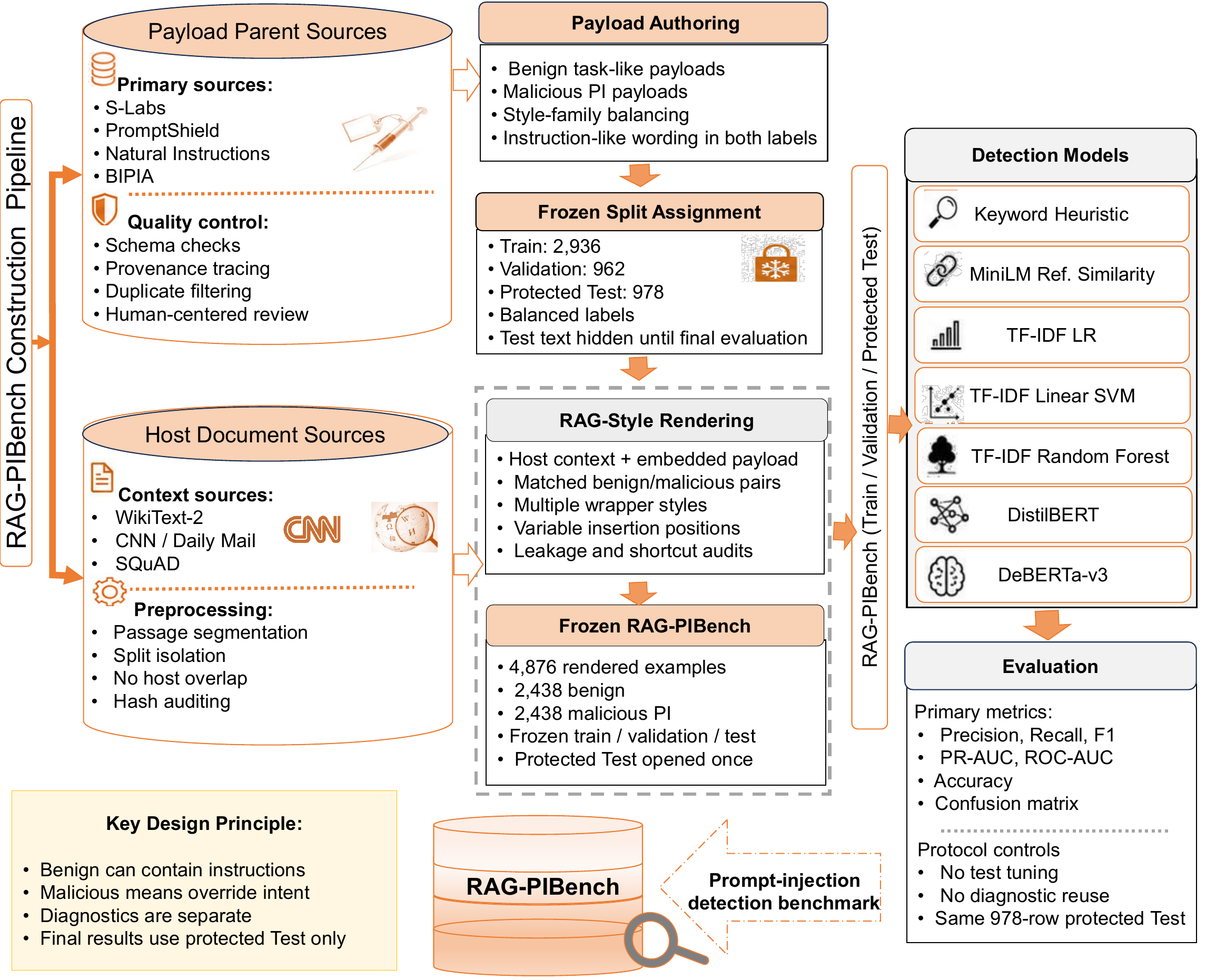}
    \caption{\ours{} construction and evaluation pipeline. Candidate host passages and payload parents are registered, filtered, and reviewed before being rendered into RAG-style contextual examples. The benchmark is frozen into train, validation, and protected-test splits with balanced benign and malicious labels.}
    \label{fig:ragshield_pipeline}
\end{figure*}

As shown in Fig.~\ref{fig:ragshield_pipeline}, \ours{} follows a staged, leakage-aware construction and evaluation workflow. The pipeline comprises four phases: (i) source registration, schema validation, and provenance tracing; (ii) candidate review and quality control; (iii) RAG-style contextual rendering and leakage auditing; and (iv) frozen split materialization, model development, and protected-test evaluation. This design maintains strict separation between dataset construction, model selection, diagnostics, and final evaluation.

\BfPara{Candidate Sources and Semantic Labels}
\label{subsec:candidate_sources_labels} 
\ours{} was constructed from host documents and embedded-content candidates that were rendered into RAG-style inputs. Each example was assigned one of two labels: \texttt{benign}, indicating ordinary instructional content that does not attempt to alter system behavior, or \path{malicious_prompt_injection}, indicating content that attempts to redirect, bypass, replace, ignore, or override the intended behavior of the assistant. A central design goal was to ensure that both classes remain task-like and instruction-bearing. The benchmark therefore focuses on distinguishing benign instructions from adversarial instruction-override attempts rather than simply detecting the presence of instructions.

\begin{figure*}[t]
\centering

    \begin{minipage}{0.47\textwidth}
    \centering
    \begin{tikzpicture}
    \begin{axis}[
        ybar,
        bar width=13pt,
        width=\linewidth,
        height=3.4cm,
        ymin=0,
        ymax=1850,
        ylabel={Selected host documents},
        symbolic x coords={CNN/DailyMail,SQuAD,WikiText-2},
        xtick=data,
        x tick label style={
            rotate=28,
            anchor=east,
            font=\small
        },
        yticklabel style={font=\small},
        ylabel style={font=\small},
        tick label style={font=\small},
        enlarge x limits=0.22,
        clip=false,
        legend style={
            at={(0.5,1.18)},
            anchor=south,
            draw=black,
            fill=white,
            font=\scriptsize,
            legend columns=1
        },
    ]
    
    \addplot[
        fill=blue!60,
        draw=black
    ] coordinates {
        (CNN/DailyMail,1500)
        (SQuAD,1500)
        (WikiText-2,500)
    };
    
    \legend{Host document sources}
    
    \node[font=\scriptsize, anchor=south]
        at ([yshift=3pt]axis cs:CNN/DailyMail,1500) {42.9\%};
    \node[font=\scriptsize, anchor=south]
        at ([yshift=3pt]axis cs:SQuAD,1500) {42.9\%};
    \node[font=\scriptsize, anchor=south]
        at ([yshift=3pt]axis cs:WikiText-2,500) {14.3\%};
    
    \end{axis}
    \end{tikzpicture}
    
    \vspace{1mm}
    \small (a) Selected host document sources
    \end{minipage}
\hfill
    \begin{minipage}{0.50\textwidth}
    \centering
    \begin{tikzpicture}
    \begin{axis}[
        ybar,
        bar width=8pt,
        width=\linewidth,
        height=3.0cm,
        ymin=0,
        ymax=92,
        ylabel={Selected payload parents},
        symbolic x coords={S-Labs,Natural Instructions,PromptShield,BIPIA},
        xtick=data,
        x tick label style={
            rotate=28,
            anchor=east,
            font=\small
        },
        yticklabel style={font=\small},
        ylabel style={font=\small},
        tick label style={font=\small},
        enlarge x limits=0.24,
        clip=false,
        legend style={
            at={(0.5,1.18)},
            anchor=south,
            draw=black,
            fill=white,
            font=\scriptsize,
            legend columns=1
        },
    ]
    
    \addplot[
        fill=blue!55,
        draw=black,
        bar shift=-5pt
    ] coordinates {
        (S-Labs,64)
        (Natural Instructions,64)
        (PromptShield,56)
        (BIPIA,8)
    };
    
    \addplot[
        fill=red!70,
        draw=black,
        bar shift=5pt
    ] coordinates {
        (S-Labs,64)
        (Natural Instructions,0)
        (PromptShield,64)
        (BIPIA,64)
    };
    
    \legend{Benign payload parents, Malicious payload parents}
    
    \node[font=\tiny, anchor=south east]
        at ([xshift=-3pt,yshift=3pt]axis cs:S-Labs,64) {33.3\%};
    
    \node[font=\tiny, anchor=south west]
        at ([xshift=3pt,yshift=9pt]axis cs:S-Labs,64) {33.3\%};
    
    \node[font=\tiny, anchor=south]
        at ([xshift=-5pt,yshift=3pt]axis cs:Natural Instructions,64) {33.3\%};
    
    \node[font=\tiny, anchor=south east]
        at ([xshift=-3pt,yshift=3pt]axis cs:PromptShield,56) {29.2\%};
    
    \node[font=\tiny, anchor=south west]
        at ([xshift=3pt,yshift=9pt]axis cs:PromptShield,64) {33.3\%};
    
    \node[font=\tiny, anchor=south]
        at ([xshift=-5pt,yshift=3pt]axis cs:BIPIA,8) {4.2\%};
    
    \node[font=\tiny, anchor=south west]
        at ([xshift=3pt,yshift=9pt]axis cs:BIPIA,64) {33.3\%};
    
    \end{axis}
    \end{tikzpicture}
    
    \vspace{1mm}
    \small (b) Primary payload-parent source composition
    \end{minipage}
\caption{Source composition of \ours{}. (a) Host-document sources used for contextual passages. (b) Payload-parent sources used to generate benign and malicious embedded content. Percentages are computed over 3,500 host documents in (a) and separately over the benign and malicious payload-parent pools (192 sources each) in (b).}
\label{fig:hardbench_source_distribution}\vspace{-5mm}
\end{figure*}
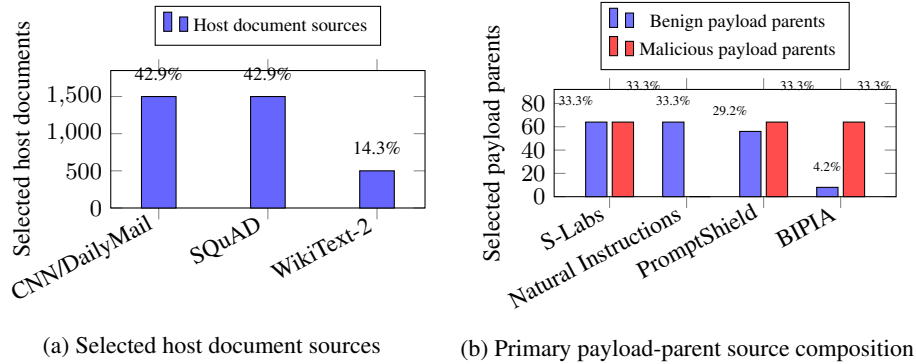

\BfPara{Schema Validation and Provenance Tracing} 
\label{subsec:schema_provenance}
Each candidate example was represented using a structured schema containing rendered text, semantic labels, identifiers, split metadata, and provenance information. Validation ensured field consistency, label correctness, and rendering compatibility. Provenance records tracked the origin and transformation of each example throughout construction, supporting auditing and reducing the risk of duplicated, unsupported, or inconsistently labeled entries.

\BfPara{Hybrid Human-Centered Review and Quality Control}
\label{subsec:human_centered_review} 
Candidate examples underwent hybrid human-centered quality control combining manual review with automated checks for schema consistency, duplication, provenance integrity, label agreement, and potential shortcut artifacts. Automated analyses were used to support review rather than assign final labels. Examples that were ambiguous, unsupported, inconsistent with the task definition, or likely to introduce artifacts were excluded or subjected to additional review. All inclusion decisions were finalized before model evaluation, and protected-test data remained inaccessible during model development.
\begin{wraptable}{r}{0.32\textwidth}
\vspace{-10mm}
\centering
\caption{Data split statistics.}
\label{tab:dataset_splits}
\footnotesize
\setlength{\tabcolsep}{3pt}

\begin{tabular}{lrrr}
\toprule
Split & Total & Benign & Mal. \\
\midrule
Train & 2936 & 1468 & 1468 \\
Valid. & 962 & 481 & 481 \\
Prot. Test & 978 & 489 & 489 \\
\midrule
Total & 4876 & 2438 & 2438 \\
\bottomrule
\end{tabular}
\vspace{-5mm}
\end{wraptable}
\BfPara{RAG-Style Contextual Rendering}
\label{subsec:rag_style_rendering} 
Retained examples were rendered into RAG-style inputs consisting of a host context and an embedded content segment. The embedded segment contains either a benign instruction or a malicious prompt-injection attempt. This design approximates realistic retrieval-augmented inputs and prevents the task from reducing to isolated prompt-fragment classification, requiring detectors to reason about instructions within their surrounding context.

\BfPara{Leakage, Shortcut, and Residual-Cue Auditing}
\label{subsec:leakage_shortcut_auditing} 
To reduce shortcut learning, the benchmark underwent leakage and residual-cue auditing prior to finalization. Audits examined duplicate content, split contamination, repeated templates, source-specific markers, formatting artifacts, and other cues that could enable non-semantic classification. Identified issues were corrected, removed, or explicitly documented. Diagnostic analyses remained separate from final model selection and protected-test evaluation.

\BfPara{Frozen Dataset Splits}
\label{subsec:frozen_splits} 
The final benchmark contains 4,876 examples divided into fixed train, validation, and protected-test splits, each balanced across benign and malicious classes. Table~\ref{tab:dataset_splits} summarizes the split statistics.

Training data were used for model fitting, validation data for hyperparameter and checkpoint selection, and the protected-test split exclusively for final evaluation. Protected-test content remained inaccessible during model development, and final results were produced through a single protected-test evaluation pass.

\subsection{Evaluation Settings}

\subsubsection{Model Families}
\label{subsec:model_families}

We evaluated two model families: baseline detectors and transformer-based classifiers. Baseline detectors provide strong lightweight reference points, while transformer models assess whether pretrained contextual representations improve prompt-injection detection under the same frozen evaluation protocol.

\vspace{-3mm}
\paragraph{Baseline Detectors} \label{subsubsec:non_transformer_baselines}
We evaluated five baseline detectors spanning rule-based, semantic-similarity, and sparse-feature approaches. \ding{172}
     \textbf{Keyword}: a deterministic detector based on lexical patterns associated with prompt injection.
    \ding{173} \textbf{MiniLM Ref.}: a semantic-similarity detector that scores inputs using maximum similarity to a malicious-reference bank.
    \ding{174} \textbf{TF-IDF LR}: logistic regression using word- and character-level TF-IDF features.
    \ding{175} \textbf{TF-IDF SVM}: a linear support-vector machine using word- and character-level TF-IDF features.
    \ding{176} \textbf{TF-IDF RF}: a random forest trained on TF-IDF representations.

These baselines provide a progression from simple lexical rules to competitive sparse-feature classifiers. Their inclusion helps determine whether transformer models provide meaningful gains beyond lightweight alternatives.

\vspace{-3mm}
\paragraph{Transformer Classifiers} \label{subsubsec:transformer_classifiers}
We evaluated two fine-tuned transformer encoder classifiers. \ding{172} \textbf{DistilBERT-base-uncased}: a compact pretrained encoder serving as an efficient contextual baseline.
    \ding{173} \textbf{DeBERTa-v3-base}: a higher-capacity encoder included to assess whether increased model complexity improves detection performance. 
    
Both models were trained on the training split, selected using validation performance, and evaluated once on the protected-test split. DistilBERT used a maximum sequence length of 256, while DeBERTa-v3 used a maximum sequence length of 192 under a conservative numerical-stability configuration. These settings were fixed prior to protected-test evaluation and were not modified afterward.

\vspace{-5mm}
\subsubsection{Training, Validation, and Protected-Test Evaluation}
\label{subsec:training_validation_test}

The experimental workflow followed a strict train/validation/protected-test protocol. Model development was confined to the training and validation splits, while the protected-test split was reserved exclusively for final evaluation. Baseline detectors were trained and tuned under this protocol, and transformer classifiers were evaluated on protected-test examples only after training and validation were complete.

To preserve evaluation integrity, train/validation transformer results were recorded separately from protected-test results, and the protected-test split was opened only for final inference. No checkpoint reselection, threshold retuning, additional training, or diagnostic feedback was permitted after protected-test exposure. The final model-comparison table was frozen only after all model families had been evaluated on the same 978-example protected-test split.

\vspace{-5mm}
\subsubsection{Separate Diagnostic Suites}
\label{subsec:diagnostic_suites}

In addition to the primary benchmark split, separate diagnostic suites were used to analyze robustness, false-positive behavior, and distributional sensitivity. These diagnostics included broad contextual examples, hard benign utility cases, and external transfer/source-audit settings.

Diagnostic suites were not used for model selection, threshold tuning, or post-hoc correction. They were maintained separately from the primary protected-test evaluation to preserve the independence of final reported results.

\vspace{-5mm}
\subsubsection{Evaluation Metrics}
\label{subsec:evaluation_metrics}

We report accuracy, precision, recall, \(F_1\), PR-AUC, ROC-AUC, and confusion-matrix counts. The malicious prompt-injection class is treated as the positive class. Let \(TP\), \(FP\), \(TN\), and \(FN\) denote true positives, false positives, true negatives, and false negatives, respectively. The operating-point metrics are defined as: Accuracy \(= \frac{TP + TN}{TP + TN + FP + FN}\), Precision \(= \frac{TP}{TP + FP}\), Recall \(= \frac{TP}{TP + FN}\), and \(F_1 = \frac{2 \cdot \text{Precision} \cdot \text{Recall}}{\text{Precision} + \text{Recall}}\). PR-AUC captures the trade-off between detecting malicious inputs and avoiding false alarms, while ROC-AUC provides a complementary threshold-independent ranking measure. We use \(F_1\) as the primary ranking metric because the benchmark is class-balanced and prompt-injection detection requires balancing missed attacks against false alarms. PR-AUC and ROC-AUC are reported to characterize ranking quality beyond the selected decision threshold.

\vspace{-5mm}
\subsubsection{Reproducibility and Artifact Registration}
\label{subsec:reproducibility}

Each stage of the \ours{} pipeline produced audit artifacts, manifests, and checksums documenting dataset lineage, split statistics, evaluation boundaries, and model outputs. The final artifact package includes the frozen protected-test comparison table, paper-ready tables and figures, split statistics, a dataset card, interpretation notes, and a reproducibility registry.

The final model-comparison table was frozen only after: (i) benchmark splits were fixed and balanced; (ii) baseline protected-test results were registered; (iii) transformer train/validation results were separated from protected-test results; (iv) transformer protected-test evaluation was completed; (v) all model families were evaluated on the same protected-test split; and (vi) no threshold retuning, diagnostic reuse, or additional training occurred during final evaluation. This workflow provides a reproducible audit trail and preserves a clear separation between model development, validation-based selection, diagnostic analysis, and final reporting.

\vspace{-5mm}
\subsubsection{Methodological Safeguards}
\label{subsec:methodological_safeguards}

Several safeguards were incorporated for evaluation validity. First, examples were retained according to benchmark policy and quality-control criteria rather than model separability. Second, the protected-test split was excluded from model selection. Third, validation-only transformer results were prevented from entering the final comparison. Fourth, diagnostic suites were separated from primary evaluation. Finally, all models were evaluated on the same protected-test split.

Together, these safeguards reduce the risk of leakage, shortcut learning, post-hoc tuning, and unfair comparison, ensuring that reported differences reflect performance under a common protected-test protocol.

\section{Results and Discussion}
\label{sec:results_discussion}

This section reports the primary protected-test results for \ours{}. All models are evaluated on the same balanced protected-test split of 978 examples (489 benign and 489 malicious), including five baseline detectors and two fine-tuned transformer classifiers. Consistent with the methodology, the protected-test split was reserved exclusively for final evaluation, with no post-test training, checkpoint reselection, threshold tuning, or diagnostic feedback. Diagnostic suites are reported separately and do not influence model ranking. Consequently, all results reflect a common protected-test comparison across detector families.

\subsection{Overall Primary Protected-Test Performance}
\label{subsec:overall_primary_test_performance}

Table~\ref{tab:primary_test_results} reports the final primary protected-test results on the balanced 978-example held-out split. Models are ordered by \(F_1\), which we use as the primary operating-point metric because prompt-injection detection requires a practical trade-off between missed malicious examples and false alarms on benign examples. Precision, recall, and \(F_1\) are computed with the malicious prompt-injection class as the positive class.

\begin{table*}[t]
\centering
\caption{Primary protected-test performance across all evaluated detectors.}
\label{tab:primary_test_results}
\footnotesize
\setlength{\tabcolsep}{4.2pt}
\renewcommand{\arraystretch}{1.08}
\begin{threeparttable}
\begin{tabular*}{\textwidth}{@{\extracolsep{\fill}}llrrrrrr@{}}
\toprule
\textbf{Class} & \textbf{Model} 
& \textbf{Prec.} & \textbf{Rec.} & \(\boldsymbol{F_1}\)
& \textbf{PR-AUC} & \textbf{ROC-AUC} & \textbf{Acc.} \\
\midrule
FT-Enc. & DistilBERT  & 0.924 & 0.869 & \textbf{0.896} & \textbf{0.968} & \textbf{0.967} & \textbf{0.899} \\
Sparse  & TF-IDF SVM  & 0.853 & 0.890 & 0.871 & 0.950 & 0.949 & 0.868 \\
Sparse  & TF-IDF LR   & 0.822 & 0.916 & 0.867 & 0.950 & 0.949 & 0.859 \\
Sparse  & TF-IDF RF   & 0.822 & 0.879 & 0.850 & 0.933 & 0.931 & 0.845 \\
SemSim  & MiniLM Ref. & 0.508 & 0.969 & 0.667 & 0.576 & 0.571 & 0.515 \\
FT-Enc. & DeBERTa-v3  & 0.500 & \textbf{1.000} & 0.667 & 0.488 & 0.487 & 0.500 \\
Rule    & Keyword     & \textbf{0.981} & 0.104 & 0.189 & 0.550 & 0.551 & 0.551 \\
\bottomrule
\end{tabular*}

\begin{tablenotes}[flushleft]
\scriptsize
\item \emph{Note:} Results are reported on the 978-example balanced protected-test split. Prec., Rec., Acc., LR, SVM, RF, FT-Enc., SemSim, and Ref. denote precision, recall, accuracy, logistic regression, support vector machine, random forest, fine-tuned encoder, semantic-similarity baseline, and reference baseline, respectively. Precision, recall, and \(F_1\) use the malicious prompt-injection class as the positive class. Bold values indicate column-best results.
\end{tablenotes}
\end{threeparttable}
\vspace{-1.5mm}
\end{table*}

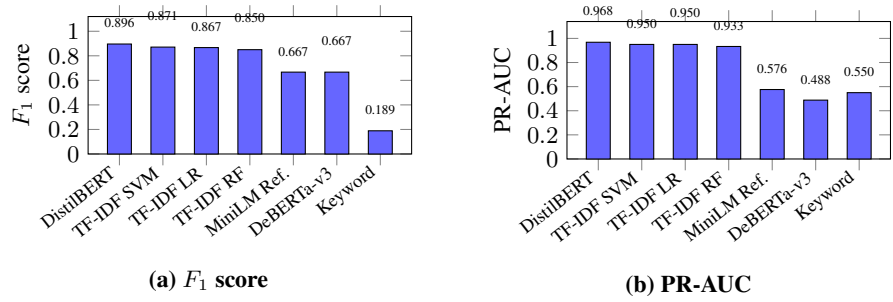
\begin{figure*}[t]
\centering

\begin{minipage}{0.48\textwidth}
\centering
\begin{tikzpicture}
\begin{axis}[
    ybar,
    bar width=9pt,
    width=\linewidth,
    height=3.4cm,
    ymin=0,
    ymax=1.12,
    ylabel={\(F_1\) score},
    symbolic x coords={
        DistilBERT,
        TF-IDF SVM,
        TF-IDF LR,
        TF-IDF RF,
        MiniLM Ref.,
        DeBERTa-v3,
        Keyword
    },
    xtick=data,
    x tick label style={
        rotate=38,
        anchor=east,
        font=\scriptsize
    },
    ytick={0,0.2,0.4,0.6,0.8,1.0},
    yticklabel style={font=\small},
    ylabel style={font=\small},
    tick label style={font=\small},
    enlarge x limits=0.12,
    clip=false
]

\addplot[
    fill=blue!60,
    draw=black
] coordinates {
    (DistilBERT,0.896)
    (TF-IDF SVM,0.871)
    (TF-IDF LR,0.867)
    (TF-IDF RF,0.850)
    (MiniLM Ref.,0.667)
    (DeBERTa-v3,0.667)
    (Keyword,0.189)
};

\node[font=\tiny, anchor=south] at ([yshift=3pt]axis cs:DistilBERT,0.896) {0.896};
\node[font=\tiny, anchor=south] at ([yshift=7pt]axis cs:TF-IDF SVM,0.871) {0.871};
\node[font=\tiny, anchor=south] at ([yshift=3pt]axis cs:TF-IDF LR,0.867) {0.867};
\node[font=\tiny, anchor=south] at ([yshift=7pt]axis cs:TF-IDF RF,0.850) {0.850};
\node[font=\tiny, anchor=south] at ([yshift=3pt]axis cs:MiniLM Ref.,0.667) {0.667};
\node[font=\tiny, anchor=south] at ([yshift=7pt]axis cs:DeBERTa-v3,0.667) {0.667};
\node[font=\tiny, anchor=south] at ([yshift=3pt]axis cs:Keyword,0.189) {0.189};

\end{axis}
\end{tikzpicture}

\vspace{1mm}
\small \textbf{(a) \(F_1\) score}
\end{minipage}
\hfill
\begin{minipage}{0.48\textwidth}
\centering
\begin{tikzpicture}
\begin{axis}[
    ybar,
    bar width=9pt,
    width=\linewidth,
    height=3.4cm,
    ymin=0,
    ymax=1.14,
    ylabel={PR-AUC},
    symbolic x coords={
        DistilBERT,
        TF-IDF SVM,
        TF-IDF LR,
        TF-IDF RF,
        MiniLM Ref.,
        DeBERTa-v3,
        Keyword
    },
    xtick=data,
    x tick label style={
        rotate=38,
        anchor=east,
        font=\scriptsize
    },
    ytick={0,0.2,0.4,0.6,0.8,1.0},
    yticklabel style={font=\small},
    ylabel style={font=\small},
    tick label style={font=\small},
    enlarge x limits=0.12,
    clip=false
]

\addplot[
    fill=blue!60,
    draw=black
] coordinates {
    (DistilBERT,0.968)
    (TF-IDF SVM,0.950)
    (TF-IDF LR,0.950)
    (TF-IDF RF,0.933)
    (MiniLM Ref.,0.576)
    (DeBERTa-v3,0.488)
    (Keyword,0.550)
};

\node[font=\tiny, anchor=south] at ([yshift=7pt]axis cs:DistilBERT,0.968) {0.968};
\node[font=\tiny, anchor=south] at ([yshift=3pt]axis cs:TF-IDF SVM,0.950) {0.950};
\node[font=\tiny, anchor=south] at ([yshift=7pt]axis cs:TF-IDF LR,0.950) {0.950};
\node[font=\tiny, anchor=south] at ([yshift=3pt]axis cs:TF-IDF RF,0.933) {0.933};
\node[font=\tiny, anchor=south] at ([yshift=3pt]axis cs:MiniLM Ref.,0.576) {0.576};
\node[font=\tiny, anchor=south] at ([yshift=3pt]axis cs:DeBERTa-v3,0.488) {0.488};
\node[font=\tiny, anchor=south] at ([yshift=3pt]axis cs:Keyword,0.550) {0.550};

\end{axis}
\end{tikzpicture}

\vspace{1mm}
\small \textbf{(b) PR-AUC}
\end{minipage}

\caption{Primary protected-test performance summary. 
Panel (a) reports the fixed operating-point \(F_1\) score, where DistilBERT achieves the strongest overall result and TF-IDF SVM provides the strongest sparse lexical baseline. 
Panel (b) reports PR-AUC, showing that strong TF-IDF baselines remain competitive for ranking malicious prompt-injection examples.}
\label{fig:primary_test_F_1_prauc}
\end{figure*}

DistilBERT achieves the strongest overall protected-test performance, with \(F_1=0.896\), precision \(=0.924\), recall \(=0.869\), PR-AUC \(=0.968\), ROC-AUC \(=0.967\), and accuracy \(=0.899\). Its confusion matrix contains 454 true negatives, 35 false positives, 64 false negatives, and 425 true positives. This indicates the best observed balance between detecting malicious prompt-injection examples and preserving benign utility. The high PR-AUC and ROC-AUC also suggest that the result is not only tied to a single decision threshold, but reflects strong ranking behavior across operating points.

The sparse lexical baselines remain highly competitive. Among them, TF-IDF SVM achieves the strongest sparse-feature baseline \(F_1\), with \(F_1=0.871\), precision \(=0.853\), recall \(=0.890\), PR-AUC \(=0.950\), and ROC-AUC \(=0.949\). TF-IDF LR is closely matched, with \(F_1=0.867\) and PR-AUC \(=0.950\). These results suggest that lexical and character-level cues capture substantial discriminative signal in \ours{}, while transformer-based contextual modeling provides the strongest overall performance.

\subsection{Transformer and Baseline Detector Comparison}
\label{subsec:transformer_nontransformer_comparison}

The protected-test results show that DistilBERT achieves the strongest overall performance, but its advantage over the best sparse baseline is modest. DistilBERT improves \(F_1\) from \(0.871\) (TF-IDF SVM) to \(0.896\), indicating that contextual representations provide additional discriminative signal while confirming that sparse lexical features capture much of the structure required for prompt-injection detection in \ours{}.

This finding has practical implications. While transformer-based detectors may be preferred when maximizing predictive performance, TF-IDF SVM and TF-IDF LR remain competitive while being less computationally expensive, easier to train, and more interpretable. Thus, prompt-injection detection studies should compare new methods against strong sparse baselines rather than only against simple heuristic detectors.

The comparison between TF-IDF SVM and TF-IDF LR also highlights a precision--recall trade-off. TF-IDF SVM achieves a slightly higher \(F_1\) score and fewer false positives (75 vs.\ 97), whereas TF-IDF LR achieves higher recall and fewer false negatives (41 vs.\ 54). These results suggest that even closely matched sparse models may support different deployment objectives, with TF-IDF SVM favoring precision and TF-IDF LR favoring recall.

\subsection{Error Trade-offs and Baseline Behavior}
\label{subsec:error_tradeoffs}

The keyword baseline serves as a lower-bound reference. It achieves very high precision (\(0.981\)) but very low recall (\(0.104\)), indicating that explicit lexical rules identify only a small subset of obvious prompt-injection attempts while missing most malicious examples. Its confusion matrix contains 488 true negatives, 1 false positive, 438 false negatives, and 51 true positives, making it unsuitable as a standalone defense despite its low false-positive rate.

The MiniLM malicious-reference baseline exhibits the opposite behavior. It achieves high recall (\(0.969\)) but low precision (\(0.508\)), producing 30 true negatives, 459 false positives, 15 false negatives, and 474 true positives. This suggests that semantic-reference methods can recover most malicious examples but may substantially over-flag benign content, creating a significant false-positive burden in practical deployments.

The TF-IDF models provide a more balanced operating point. TF-IDF SVM, TF-IDF logistic regression, and TF-IDF random forest all achieve \(F_1\) scores above \(0.84\), indicating that prompt-injection detection benefits from richer lexical and structural representations than simple keyword matching. The strong performance of these sparse-feature models suggests that much of the discriminative signal in \ours{} can be captured through word- and character-level patterns while maintaining a substantially better precision--recall balance than either the keyword or semantic-reference baselines.

\subsection{Transformer-Specific Findings}
\label{subsec:transformer_specific_findings}

The transformer results reveal a sharp contrast between DistilBERT and DeBERTa-v3 under the frozen evaluation protocol. DistilBERT achieves the strongest overall protected-test performance, whereas DeBERTa-v3 exhibits a degenerate all-positive prediction pattern, classifying every example as malicious. As a result, DeBERTa-v3 achieves perfect recall (\(1.000\)) but only \(0.500\) precision, with 489 false positives and no true negatives. Its ROC-AUC and PR-AUC also fall below those of the strongest baseline models, indicating limited ranking ability under the evaluated configuration.

This outcome should not be interpreted as evidence that DeBERTa-v3 is inherently unsuitable for prompt-injection detection. Rather, it reflects the specific conservative training and numerical-stability configuration used in this study. While the final model completed training and evaluation successfully, it failed to learn a useful decision boundary on the protected-test split. The result highlights that increased model capacity does not necessarily translate into improved detection performance under a frozen, leakage-aware evaluation protocol. More broadly, the contrast between DistilBERT and DeBERTa-v3 underscores the importance of transparent reporting. Reporting only successful transformer results would overstate the reliability of transformer-based detection. Including unsuccessful outcomes provides a more complete characterization of model behavior and strengthens the credibility of the evaluation.

\subsection{Ranking Metrics and Operating-Point Metrics}
\label{subsec:ranking_operating_metrics}

The results illustrate the value of reporting both operating-point and ranking metrics. Accuracy, Precision, Recall, and \(F_1\) characterize behavior at the selected decision threshold, whereas PR-AUC and ROC-AUC measure ranking quality across thresholds. DistilBERT performs strongly under both views, achieving the highest \(F_1\), PR-AUC, and ROC-AUC, indicating that its advantage is not tied to a particular operating point.

The sparse TF-IDF models also demonstrate strong ranking performance. TF-IDF logistic regression and TF-IDF SVM both achieve PR-AUC values near \(0.95\) and ROC-AUC values near \(0.95\), showing that lightweight sparse-feature models can effectively rank malicious examples despite trailing DistilBERT on overall protected-test performance. This finding is particularly relevant for deployments where computational efficiency or model simplicity is important. In contrast, DeBERTa-v3 achieves perfect recall at the selected threshold but poor ranking performance. This result highlights that recall alone can be misleading: a detector that labels every example as malicious will never miss an attack, but it will also reject all benign content. Effective prompt-injection detection therefore requires both strong ranking ability and a practical operating point that balances attack detection against false alarms.

\subsection{Implications for RAG Prompt-Injection Detection}
\label{subsec:rag_detection_implications}

The results have several implications for RAG prompt-injection detection. First, contextual transformer representations can improve performance, as evidenced by DistilBERT achieving the strongest overall protected-test result. This suggests that pretrained contextual encoders provide additional signal for distinguishing malicious instruction-override attempts from benign embedded instructions.

Second, strong baseline detectors remain essential. TF-IDF SVM and TF-IDF logistic regression perform competitively with DistilBERT and substantially outperform the simpler keyword and semantic-reference baselines. Consequently, future studies should compare proposed methods against strong sparse-feature models rather than only against heuristic detectors.

Third, false-positive control is as important as attack coverage. Both the MiniLM malicious-reference baseline and DeBERTa-v3 achieve very high recall, but only at the cost of excessive false positives. In practical RAG deployments, over-blocking benign content can reduce system utility, disrupt legitimate workflows, and increase human-review burden. Effective detectors must therefore balance precision and recall rather than optimizing recall alone.

Finally, benchmark construction and evaluation methodology play a critical role in the credibility of reported results. By employing a frozen train/validation/protected-test protocol, separating diagnostic analyses from final model ranking, and preventing validation-only results from entering the primary comparison, \ours{} reduces opportunities for leakage and evaluation bias, yielding a more reliable assessment of prompt-injection detection performance.

\subsection{Discussion of Benchmark Characteristics}
\label{subsec:benchmark_difficulty}

\ours{} is designed to evaluate prompt-injection detection in contextualized inputs rather than isolated attack strings. The strong performance of the TF-IDF models indicates that the task contains learnable lexical and structural signals, while the large gap between keyword heuristics and learned sparse models shows that these signals extend beyond simple rule-based matching.

The modest improvement of DistilBERT over the strongest TF-IDF baselines suggests that contextual semantics provide additional discriminative information, but not enough to render sparse-feature models ineffective. This is a desirable property of the benchmark: it is neither so simple that keyword rules solve the task nor so dependent on large neural models that classical baselines become uninformative.

The DeBERTa-v3 results further indicate that model capacity alone does not determine performance. Training stability, optimization behavior, sequence-length constraints, and checkpoint selection can substantially affect protected-test outcomes. This observation reinforces the importance of reproducible evaluation protocols and transparent reporting of both successful and unsuccessful model behavior.

\subsection{Limitations}
\label{subsec:limitations}

Several limitations should be acknowledged. First, \ours{} evaluates binary prompt-injection detection rather than downstream mitigation. Practical RAG systems may combine detection with policies for isolating, sanitizing, ignoring, quoting, or escalating retrieved content, which are outside the scope of this study.

Second, although the benchmark incorporates style matching, provenance tracing, duplicate checks, and leakage auditing, no benchmark can eliminate all distributional artifacts. Accordingly, the reported results should be interpreted as performance on \ours{} rather than as evidence of universal robustness to prompt-injection attacks.

Third, the transformer evaluation is limited to two encoder-based models under a frozen training and selection protocol. Broader coverage, including larger encoders, decoder-only models, instruction-tuned classifiers, and retrieval-aware architectures, may reveal additional performance characteristics. The DeBERTa-v3 findings should therefore be interpreted within the specific training and numerical-stability configuration evaluated in this work.

Fourth, the primary protected-test split is class-balanced. While this design supports fair model comparison and straightforward interpretation of \(F_1\), real deployments may exhibit substantially different class priors. In such settings, calibration and threshold selection may require additional validation procedures that remain independent of protected-test data.

Finally, the benchmark is limited to English textual inputs. Future work should examine multilingual prompt-injection attacks, cross-domain transfer, adversarially generated content, and application-specific RAG environments.

\subsection{Summary of Findings}
\label{subsec:summary_findings}

Overall, the final protected-test results show that \ours{} provides an informative evaluation setting for prompt-injection detection. DistilBERT achieves the strongest overall performance, with the best \(F_1\), PR-AUC, ROC-AUC, and accuracy. However, strong TF-IDF baselines remain highly competitive, demonstrating that lightweight models can capture substantial prompt-injection signals. Keyword rules are too brittle because they miss most malicious examples, while semantic-reference detection and DeBERTa-v3 show that high recall can be misleading when false positives are uncontrolled. These findings support three main conclusions. First, prompt-injection detection should be evaluated against strong baseline detectors, not only weak heuristics. Second, protected-test evaluation must be separated from validation and diagnostic analysis to avoid inflated claims. Third, transparent reporting of both successful and failed model behavior is necessary for reliable progress in RAG security evaluation.

\section{Concluding Remarks}
\label{sec:conclusion}

This paper introduced \ours{}, a leakage-aware benchmark for binary prompt-injection detection in RAG-style contextual inputs. The benchmark was designed to evaluate whether detectors can distinguish benign embedded instructions from malicious instruction-override attempts under a frozen train, validation, and protected-test protocol. To support reproducible evaluation, the construction process incorporated provenance tracing, schema validation, hybrid human-centered quality control, contextual rendering, shortcut auditing, and strict split isolation. The final evaluation compared heuristic, semantic-reference, sparse-feature, and transformer-based detectors on a common 978-example protected-test split.

The results demonstrate that contextual transformer representations can improve prompt-injection detection performance. DistilBERT achieved the strongest overall protected-test performance, obtaining an \(F_1\) score of \(0.896\), PR-AUC of \(0.968\), and ROC-AUC of \(0.967\). At the same time, strong sparse-feature baselines remained highly competitive, with TF-IDF SVM achieving an \(F_1\) score of \(0.871\) and TF-IDF logistic regression achieving a PR-AUC of \(0.950\). These findings indicate that sparse lexical and character-level representations capture substantial prompt-injection signal and should remain part of future benchmark comparisons.

The evaluation also revealed important failure modes. Keyword heuristics achieved high precision but very low recall, demonstrating the limitations of explicit trigger-based rules. The semantic-reference baseline achieved high recall but generated many false positives, highlighting the risk of over-flagging benign instructional content. Under the evaluated configuration, DeBERTa-v3 collapsed to an all-positive prediction pattern, illustrating that larger model capacity does not necessarily translate into better prompt-injection detection performance.

Overall, the results suggest that effective prompt-injection detection requires both strong benchmark design and rigorous evaluation methodology. They further demonstrate the value of comparing transformer models against competitive lightweight baselines and of maintaining strict separation between model development, diagnostic analysis, and final testing. Future work should expand benchmark coverage through broader model families, multilingual and cross-domain evaluation, adversarially generated attacks, multimodal prompt-injection scenarios, and deployment-oriented studies that jointly evaluate detection and mitigation strategies.

\appendix

\section{Supplementary Reproducibility and Evaluation Details}
\label{app:reproducibility_evaluation_details}

This appendix provides supplementary details that support reproducibility and clarify the evaluation boundaries of \ours{}. It does not introduce additional model-selection criteria and does not alter the primary protected-test results reported in the main paper.

\subsection{Dataset Construction and Label Policy}
\label{app:label_policy}

\ours{} is formulated as a binary prompt-injection detection benchmark for RAG-style contextual inputs. Each example contains a rendered host context and an embedded content segment. The negative class, \texttt{benign}, denotes embedded content that may contain a legitimate task, request, instruction, or text segment but does not attempt to override the surrounding task or system behavior. The positive class, \path{malicious_prompt_injection}, denotes embedded content that attempts to override, ignore, redirect, bypass, replace, or manipulate the intended instruction-following behavior of the downstream assistant.

A key design principle is that both classes may contain instruction-like language. Therefore, the benchmark is not a simple test of whether imperative wording appears in the input. Instead, the task is to distinguish benign embedded instructions from malicious instruction-override behavior. Candidate rows that were ambiguous, unsupported, duplicated, inconsistent with the task definition, or likely to introduce shortcut artifacts were excluded, quarantined, or sent to additional confirmation before final split materialization.

The final benchmark contains 4,876 examples across frozen train, validation, and protected-test splits. The training split contains 2,936 examples, the validation split contains 962 examples, and the protected-test split contains 978 examples. Each split is balanced between benign and malicious prompt-injection examples.

\subsection{Reproducibility Phase Registry}
\label{app:phase_registry}

Table~\ref{tab:app_phase_registry} summarizes the registered stages used to separate dataset construction, model selection, protected-test access, and final artifact generation.

\begin{table}[!htbp]
\centering
\caption{Reproducibility controls for the final \ours{} results.}
\label{tab:app_phase_registry}
\scriptsize
\setlength{\tabcolsep}{3.5pt}
\renewcommand{\arraystretch}{1.06}
\begin{threeparttable}
\begin{tabularx}{\linewidth}{@{}p{0.24\linewidth}Yp{0.20\linewidth}@{}}
\toprule
\textbf{Stage} & \textbf{Reproducibility control} & \textbf{Protocol ID} \\
\midrule
Data freeze
& Fixed the train, validation, and protected-test rows before model development.
& 0R \\

Sparse/rule baselines
& Registered the final protected-test results for sparse and rule-based baselines.
& 0AA.2G/0AA.2H \\

Transformer training
& Restricted training to the train split and checkpoint selection to validation only.
& 0AA.2L.1-R4 \\

Validation gate
& Prevented validation-only transformer results from entering the final test table.
& 0AA.2M-LITE \\

Protected-test transfer
& Opened the protected-test split only for final transformer inference.
& 0AA.2N.0 \\

Transformer evaluation
& Produced the final transformer protected-test metrics.
& 0AA.2N.1 \\

Artifact freeze
& Froze the final tables, figures, dataset card, and paper artifacts.
& 0AA.2O-R1 \\
\bottomrule
\end{tabularx}

\begin{tablenotes}[flushleft]
\scriptsize
\item \emph{Note:} The registry separates data freezing, model selection, protected-test access, and final artifact generation.
\end{tablenotes}
\end{threeparttable}
\vspace{-2mm}
\end{table}

This staged workflow ensures that validation results, diagnostic analyses, and protected-test results remain separated. The protected-test split was not used for training, threshold retuning, or checkpoint reselection.

\subsection{Leakage and Shortcut Audit Summary}
\label{app:leakage_shortcut_audit}

Table~\ref{tab:app_leakage_audit} summarizes the main leakage and shortcut-control checks used during benchmark construction and evaluation. These checks clarify the leakage-aware protocol. They do not imply that all possible distributional artifacts are eliminated, but they document the controls used to reduce split contamination, duplicate reuse, and post-hoc evaluation leakage.

\begin{table}[!htbp]
\centering
\caption{Leakage and shortcut-control audits used in \ours{}.}
\label{tab:app_leakage_audit}
\scriptsize
\setlength{\tabcolsep}{3.2pt}
\renewcommand{\arraystretch}{1.04}
\begin{threeparttable}
\begin{tabularx}{\linewidth}{@{}p{0.23\linewidth}YY@{}}
\toprule
\textbf{Audit} & \textbf{Check} & \textbf{Registered outcome} \\
\midrule
Rendered-text duplicates
& Exact and hash-level duplicates were checked across train, validation, and protected-test splits.
& No cross-split rendered-text duplicate leakage was registered. \\

Host-context isolation
& Host documents and segmented passages were assigned before final rendering to reduce cross-split overlap.
& Host overlap was controlled through split isolation and audit checks. \\

Payload-parent tracking
& Payload-parent sources and semantic families were tracked during candidate selection and rendering.
& Payload-parent usage was registered and controlled during benchmark construction. \\

Schema and provenance validation
& Required fields, labels, split metadata, row identifiers, and provenance fields were validated.
& Invalid, unsupported, or inconsistent rows were excluded, quarantined, or sent for further confirmation. \\

Shortcut and residual-cue review
& Formatting artifacts, source-specific cues, repeated templates, and residual shortcut indicators were reviewed.
& Problematic rows or patterns were corrected, excluded, quarantined, or explicitly registered. \\

Protected-test access boundary
& Exposure of protected-test text during train/validation model development was checked.
& Protected-test text was withheld during train/validation phases and opened only for final evaluation. \\

Diagnostic separation
& Use of diagnostic suites for model selection, threshold tuning, or post-hoc correction was checked.
& Diagnostic suites were kept separate from final model ranking and protected-test decisions. \\
\bottomrule
\end{tabularx}

\begin{tablenotes}[flushleft]
\scriptsize
\item \emph{Note:} This table summarizes registered construction and evaluation safeguards. These controls support leakage-aware benchmark construction, but do not imply that any finite benchmark is free of all possible residual distributional cues.
\end{tablenotes}
\end{threeparttable}
\vspace{-2mm}
\end{table}

\subsection{Confusion Matrices on the Primary Protected-Test Split}
\label{app:confusion_matrices}

The main paper reports the full metric table. To make the operating-point behavior transparent, Table~\ref{tab:app_confusion_matrices} reports the corresponding confusion matrices on the same 978-example primary protected-test split. The positive class is \path{malicious_prompt_injection}.

\begin{table}[!htbp]
\centering
\caption{Protected-test confusion matrices for all evaluated detectors.}
\label{tab:app_confusion_matrices}
\footnotesize
\setlength{\tabcolsep}{8pt}
\renewcommand{\arraystretch}{1.05}
\begin{threeparttable}
\begin{tabular}{@{}lrrrr@{}}
\toprule
\textbf{Model} & \textbf{TN} & \textbf{FP} & \textbf{FN} & \textbf{TP} \\
\midrule
DistilBERT  & 454 &  35 &  64 & 425 \\
TF-IDF SVM  & 414 &  75 &  54 & 435 \\
TF-IDF LR   & 392 &  97 &  41 & 448 \\
TF-IDF RF   & 396 &  93 &  59 & 430 \\
MiniLM Ref. &  30 & 459 &  15 & 474 \\
DeBERTa-v3  &   0 & 489 &   0 & 489 \\
Keyword     & 488 &   1 & 438 &  51 \\
\bottomrule
\end{tabular}

\begin{tablenotes}[flushleft]
\scriptsize
\item \emph{Note:} TN, FP, FN, and TP denote true negatives, false positives, false negatives, and true positives. The malicious prompt-injection class is treated as the positive class. Model abbreviations follow Table~\ref{tab:primary_test_results}.
\end{tablenotes}
\end{threeparttable}
\vspace{-2mm}
\end{table}

These counts show why aggregate metrics alone are insufficient. For example, the keyword heuristic has very few false positives but misses most malicious examples, while the MiniLM reference baseline and DeBERTa-v3 substantially over-flag benign examples.

\subsection{Separate Diagnostic Suites}
\label{app:diagnostic_suites}

Diagnostic suites were used only for behavioral analysis. They were not used for final model ranking, threshold tuning, checkpoint selection, or post-hoc correction of primary protected-test results. Table~\ref{tab:app_diagnostic_suites} summarizes their role.

\begin{table}[!htbp]
\centering
\caption{Diagnostic suites and evaluation boundaries.}
\label{tab:app_diagnostic_suites}
\footnotesize
\setlength{\tabcolsep}{4pt}
\renewcommand{\arraystretch}{1.06}
\begin{threeparttable}
\begin{tabularx}{\linewidth}{@{}p{0.30\linewidth}YY@{}}
\toprule
\textbf{Suite} & \textbf{Purpose} & \textbf{Boundary} \\
\midrule
Contextual diagnostic
& Rendered-distribution behavior.
& Reporting only; not used for final ranking. \\

Hard benign diagnostic
& Stress test for benign false positives.
& Reporting only; not used for threshold tuning. \\

External transfer audit
& Distribution-sensitivity analysis.
& Reporting only; not merged into the primary evaluation. \\
\bottomrule
\end{tabularx}

\begin{tablenotes}[flushleft]
\scriptsize
\item \emph{Note:} Diagnostic suites are excluded from model selection, threshold tuning, and primary protected-test ranking.
\end{tablenotes}
\end{threeparttable}
\vspace{-2mm}
\end{table}

This separation allows diagnostic suites to reveal model weaknesses without compromising the independence of the protected-test evaluation.

\subsection{Model Configuration Summary}
\label{app:model_configuration}

Table~\ref{tab:app_model_config} summarizes the model families and the main evaluation boundary for each group. Detailed metric values are reported in the main Results section.

\begin{table}[!htbp]
\centering
\caption{Model configuration and evaluation boundaries.}
\label{tab:app_model_config}
\scriptsize
\setlength{\tabcolsep}{3.5pt}
\renewcommand{\arraystretch}{1.06}
\begin{threeparttable}
\begin{tabularx}{\linewidth}{@{}p{0.23\linewidth}YY@{}}
\toprule
\textbf{Detector} & \textbf{Configuration} & \textbf{Evaluation boundary} \\
\midrule
Keyword
& Deterministic lexical-pattern heuristic.
& No train-time tuning. \\

MiniLM Ref.
& Maximum similarity to the malicious-reference bank.
& Decision threshold selected on validation only. \\

TF-IDF LR
& Word- and character-level TF-IDF features.
& Hyperparameters selected on validation only. \\

TF-IDF SVM
& Word- and character-level TF-IDF features.
& Hyperparameters selected on validation only. \\

TF-IDF RF
& Word-level TF-IDF features.
& Hyperparameters selected on validation only. \\

DistilBERT
& Transformer encoder classifier with maximum sequence length 256.
& Trained on the train split, selected on validation, and evaluated once on the protected-test split. \\

DeBERTa-v3
& Transformer encoder classifier with maximum sequence length 192 under a conservative stability configuration.
& Trained on the train split, selected on validation, and evaluated once on the protected-test split. \\
\bottomrule
\end{tabularx}

\begin{tablenotes}[flushleft]
\scriptsize
\item \emph{Note:} LR, SVM, RF, and Ref. denote logistic regression, support vector machine, random forest, and reference baseline, respectively. Detector names follow Table~\ref{tab:primary_test_results}. The protected-test split is used only for final evaluation.
\end{tablenotes}
\end{threeparttable}
\vspace{-2mm}
\end{table}

The final evaluation boundary is fixed as follows: training uses only the training split, model and checkpoint selection use only the validation split, and the protected-test split is used only once for final evaluation after configuration freeze. Diagnostic suites are excluded from final model ranking and are reported only as supplementary behavioral evidence.


\begin{thebibliography}{10}
\providecommand{\url}[1]{\texttt{#1}}
\providecommand{\urlprefix}{URL }
\providecommand{\doi}[1]{https://doi.org/#1}

\bibitem{choi2025attributing}
Choi, S., Mohaisen, D.: Attributing {ChatGPT}-generated source codes. IEEE
  Transactions on Dependable and Secure Computing  \textbf{22}(4),  3602--3615
  (2025). \doi{10.1109/TDSC.2025.3535218}

\bibitem{destefano2024rag}
De~Stefano, G., Sch{\"o}nherr, L., Pellegrino, G.: Rag and roll: An end-to-end
  evaluation of indirect prompt manipulations in {LLM}-based application
  frameworks (2024)

\bibitem{gallifant2025sparse}
Gallifant, J., Chen, S., Sasse, K., Aerts, H., Hartvigsen, T., Bitterman, D.:
  Sparse autoencoder features for classifications and transferability. In:
  Proceedings of the 2025 Conference on Empirical Methods in Natural Language
  Processing. pp. 29927--29951 (2025)

\bibitem{geng2026prompt}
Geng, T., Xu, Z., Qu, Y., Wong, W.E.: Prompt injection attacks on large
  language models: A survey of attack methods, root causes, and defense
  strategies. Computers, Materials \& Continua  \textbf{87}(1), ~4 (2026).
  \doi{10.32604/cmc.2025.074081}

\bibitem{guan2025privacy}
Guan, S., Kwok, H.C., Law, N.F., Stiglic, G., Qin, H., Hui, V.: Privacy
  challenges and solutions in retrieval-augmented generation-enhanced {LLM}s
  for healthcare chatbots: A review of applications, risks, and future
  directions. arXiv preprint arXiv:2511.11347  (2025)

\bibitem{hadiprakoso2025adaptive}
Hadiprakoso, R.B., Wilujengning, W., Amiruddin, A.: Adaptive multi-layer
  framework for detecting and mitigating prompt injection attacks in large
  language models. J. Inf. Syst. Eng. Bus. Intell  \textbf{11},  473--487
  (2025)

\bibitem{helbling2023llm}
Helbling, A., Phute, M., Hull, M., Chau, D.H.: Llm self defense: By self
  examination, llms know they are being tricked. arXiv e-prints pp. arXiv--2308
  (2023)

\bibitem{jacob2025promptshield}
Jacob, D., Alzahrani, H., Hu, Z., Alomair, B., Wagner, D.: Promptshield:
  Deployable detection for prompt injection attacks. In: Proceedings of the
  15th ACM Conference on Data and Application Security and Privacy. pp.
  341--352. CODASPY '25, Association for Computing Machinery (2025).
  \doi{10.1145/3714393.3726501}

\bibitem{jaffal2025large}
Jaffal, N.O., Alkhanafseh, M., Mohaisen, D.: Large language models in
  cybersecurity: A survey of applications, vulnerabilities, and defense
  techniques. AI  \textbf{6}(9), ~216 (2025). \doi{10.3390/ai6090216}

\bibitem{kharma2026security}
Kharma, M.F., Choi, S., Alkhanafseh, M., Mohaisen, D.: Security and quality in
  {LLM}-generated code: A multi-language, multi-model analysis. IEEE
  Transactions on Dependable and Secure Computing  (2026)

\bibitem{kokkula2024palisade}
Kokkula, S., Somanathan, R., Nandavardhan, R., Aashishkumar, Divya, G.:
  Palisade -- prompt injection detection framework. arXiv preprint
  arXiv:2410.21146  (2024). \doi{10.48550/arXiv.2410.21146}

\bibitem{latibari2024transformers}
Latibari, B.S., Nazari, N., Chowdhury, M.A., Gubbi, K.I., Fang, C., Ghimire,
  S., Hosseini, E., Sayadi, H., Homayoun, H., Salehi, S., et~al.: Transformers:
  A security perspective. IEEE Access  \textbf{12},  181071--181105 (2024)

\bibitem{lin2025large}
Lin, J., Mohaisen, D.: From large to mammoth: A comparative evaluation of large
  language models in vulnerability detection. In: Proceedings of the 32nd
  Annual Network and Distributed System Security Symposium (NDSS). Internet
  Society, San Diego, CA, USA (2025)

\bibitem{liu2025wainjectbench}
Liu, Y., Xu, R., Wang, X., Jia, Y., Gong, N.Z.: {WAInjectBench}: Benchmarking
  prompt injection detections for web agents. arXiv preprint arXiv:2510.01354
  (2025). \doi{10.48550/arXiv.2510.01354}

\bibitem{ogdu2025adaptive}
{\"O}{\u{g}}d{\"u}, {\c{C}}.U., Arslano{\u{g}}lu, K., Karak{\"o}se, M.: An
  adaptive multi-agent llm-based clinical decision support system integrating
  biomedical rag and web intelligence. IEEE Access  \textbf{13},
  167390--167404 (2025). \doi{10.1109/ACCESS.2025.3613340}

\bibitem{oro2024evaluating}
Oro, E., Granata, F.M., Lanza, A., Bachir, A., De~Grandis, L., Ruffolo, M.:
  Evaluating retrieval-augmented generation for question answering with large
  language models. In: Proceedings of the 4th National Conference on Artificial
  Intelligence (Ital-IA 2024). pp. 12--17. Naples, Italy (2024)

\bibitem{rakin2024leveraging}
Rakin, S., Shibly, M.A., Hossain, Z.M., Khan, Z., Akbar, M.M.: Leveraging the
  domain adaptation of retrieval augmented generation models for question
  answering and reducing hallucination. arXiv preprint arXiv:2410.17783  (2024)

\bibitem{shrivastav2025effective}
Shrivastav, V.A.: An Effective Approach to Protecting Large Language Model
  Applications From Prompt Injection Attacks in E-Commerce. Master's thesis,
  Arizona State University (2025)

\bibitem{si2025seconrag}
Si, X., Zhu, M., Qin, S., Yu, L., Zhang, L., Liu, S., Li, X., Duan, R., Liu,
  Y., Jia, X.: {SeCon-RAG}: A two-stage semantic filtering and conflict-free
  framework for trustworthy rag. In: Advances in Neural Information Processing
  Systems (2025). \doi{10.48550/arXiv.2510.09710}, neurIPS 2025 poster;
  arXiv:2510.09710

\bibitem{wang2025promptsleuth}
Wang, M., Zhang, Y., Gu, G.: Promptsleuth: Detecting prompt injection via
  semantic intent invariance  (2025)

\bibitem{watsonbenjamin2025contextfusion}
Watson~Benjamin, M.I.S., K, A.J.V., Yadav, R., Reddy~Thippareddy, V.S., Raju,
  B.: Context-fusion: An intelligent retrieval-augmented conversational ai
  framework with multi-model support. In: 2025 3rd International Conference on
  Self Sustainable Artificial Intelligence Systems (ICSSAS). pp. 965--971
  (2025). \doi{10.1109/ICSSAS66150.2025.11081255}

\bibitem{yi2025bipia}
Yi, J., Xie, Y., Zhu, B., Kiciman, E., Sun, G., Xie, X., Wu, F.: Benchmarking
  and defending against indirect prompt injection attacks on large language
  models. In: Proceedings of the 31st ACM SIGKDD Conference on Knowledge
  Discovery and Data Mining. pp. 1809--1820. KDD '25, Association for Computing
  Machinery (2025). \doi{10.1145/3690624.3709179}

\bibitem{zhan2024injecagent}
Zhan, Q., Liang, Z., Ying, Z., Kang, D.: Injecagent: Benchmarking indirect
  prompt injections in tool-integrated large language model agents. In:
  Findings of the Association for Computational Linguistics: ACL 2024. pp.
  10471--10506 (2024)

\bibitem{zhou2025trustrag}
Zhou, H., Lee, K.H., Zhan, Z., Chen, Y., Li, Z., Wang, Z., Haddadi, H., Yilmaz,
  E.: Trustrag: Enhancing robustness and trustworthiness in retrieval-augmented
  generation. arXiv preprint arXiv:2501.00879  (2025)

\end{thebibliography}
\end{document}